\documentclass[%
 aip,
 apl,%
 amsmath,amssymb,
 reprint,%
]{revtex4-1}

\usepackage{graphicx}
\usepackage{dcolumn}
\usepackage{bm}

\begin{document}

\title[Nonlinear modeling of magnetic nanoparticles]{Nonlinear simulations to optimize magnetic nanoparticle hyperthermia}

\author{Daniel B. Reeves}
 \affiliation{Department of physics and astronomy, Dartmouth College, Hanover, NH 03755, USA}
 \email{dbr@Dartmouth.edu.}
 
\author{John B. Weaver}%
 \affiliation{Department of physics and astronomy, Dartmouth College, Hanover, NH 03755, USA}
 \altaffiliation[]{Radiology Department, Geisel School of Medicine}

\date{\today}
\begin{abstract}

Magnetic nanoparticle hyperthermia is an attractive emerging cancer treatment, but the acting microscopic energy deposition mechanisms are not well understood and optimization suffers. We describe several approximate forms for the characteristic time of N\'{e}el rotations with varying properties and external influences. We then present stochastic simulations that show agreement between the approximate expressions and the micromagnetic model. The simulations show nonlinear imaginary responses and associated relaxational hysteresis due to the field and frequency dependencies of the magnetization. This suggests efficient heating is possible by matching fields to particles instead of resorting to maximizing the power of the applied magnetic fields. 
\end{abstract}

\maketitle

Magnetic nanoparticle (MNP) hyperthermia is considered a potentially useful addition to current cancer treatment modalities \cite{PANK} yet consensus has not been reached as to the precise mechanism of nanoparticle heating \cite{MAMI}. Simple models including linear response \cite{HERGT} and Stoner-Wohlfarth hysteresis \cite{STOWO} have been used to predict hyperthermia performance from various MNPs. Both models are approximations requiring small applied fields and equilibrium respectively, but occasionally have been applied beyond their valid range to predict optimal heating parameters. This is confusing to readers not familiar with the theory, and data are often in conflict with the theoretical predictions \cite{HERGT,MEH}.

In response to this, we demonstrate a more general approach to modeling MNP heating using nonlinear stochastic differential equations. We examine the phenomena of field dependent characteristic timescales, relaxational hysteresis curves, and nonlinear imaginary magnetization responses with the hope of informing decisions to optimize hyperthermia.

A N\'{e}el rotation model for MNP hyperthermia is justified because single-domain ferromagnetic particles experimentally display the best heating properties to date \cite{MEH}. Biological targeting schemes may also direct decisions for particle sizes, shapes, or surface construction \cite{OMID} but these choices are beyond the scope of this paper. An upper bound on the field's power could be the limit where hyperthermia's benefit of specific cytotoxicity is overwhelmed by eddy-current damage of healthy tissue. In experiment\cite{BREZ}, patients could tolerate fields with a product of field amplitude and frequency below $10^4$T/s so we do not exceed this value.

The time dynamics of MNPs can be calculated using the micromagnetic stochastic differential equation of Landau, Lifshitz, and Gilbert. The `LLG' equation derives from the Larmor precession of a spin in a magnetic field appended by a phenomenological velocity-dependent damping term \cite{GILB}. It is written in terms of the normalized magnetic moment direction $\mathbf{m}$ of each nanoparticle and the effective field $\mathbf{H}$ that can be defined as a partial derivative of the free energy with respect to the magnetic moment. The LLG equation is then for each particle (denoted with subscript $i$)
\begin{equation} \frac{\mathrm{d}\mathbf{m}_i}{\mathrm{d}t}= \frac{\gamma}{1+\alpha^2}\left[\mathbf{H} \times \mathbf{m}_i+\alpha\mathbf{m}_i \times \left(\mathbf{H} \times\mathbf{m}_i\right) \right] \end{equation}
with the electron gyromagnetic ratio $\gamma=1.76\cdot10^{11}$Hz/T, and a dimensionless magnetic damping parameter $\alpha$. We include additional physics including anisotropy, dipole-dipole interactions, and thermal fluctuations by modifying the free energy and thus the effective field. Now,
 \begin{equation} \mathbf{H}=\underbrace{H_o\hat{z}\cos{\omega t}}_\mathrm{applied}+\underbrace{\frac{2\hat{n}E_k}{\mu}\mathbf{m}_i\cdot\hat{n}_i}_\mathrm{anisotropy}+\underbrace{\frac{\mu\mu_0}{4\pi d^3}\langle\mathbf{m}\rangle}_\mathrm{dipole}+\underbrace{\mathbf{h}(t)}_\mathrm{stochastic} \label{heff}\end{equation}
where the externally applied field has amplitude $H_o$ and frequency $\omega=2\pi f$. We assume a single anisotropy axis $\hat{n}_i$ in a random direction for each particle arising from shape and crystallinity effects; this creates two energy minima, and requires a rotation of the particle to overcome the anisotropy energy barrier $E_K=K_aV_c$ with anisotropy constant $K_a$ and core volume $V_c$ assumed the same for all particles (a size distribution will change the dynamics, but presently we avoid this complication). The magnitude of the moment is $\mu=M_sV_c$ with saturation magnetization $M_s$. A mean dipole field is included that depends on the average magnetization of all the other particles; its strength determined from the magnetic moment, the permeability of free space $\mu_0$, and an average particle spacing $d= {c_N}^{-3}$ that is computed from the particle concentration $c_N$.
 
Thermal fluctuations of the field are included with a stochastic field with zero mean and unit standard deviation 
\begin{equation}\langle \mathbf{h}(t)\rangle=0, \hspace{2mm} \langle\mathbf{h}_j(t)\mathbf{h}_k(t')\rangle=\frac{2kT\alpha}{\mu\gamma}\delta_{jk}\delta(t-t')\label{delta}\end{equation}
where the Dirac delta function implies the noise field is white and is spatially correlated by the Kronecker delta where the indices imply the direction e.g., $j,k\in x,y,z$.  
 
Before solving numerically, it is possible to glean some insight with analytical approximations. The timescale of a thermal rotation over the anisotropy barrier is referred to as the N\'{e}el relaxation time\cite{FANNIN}
\begin{equation} \tau_{_\mathrm{N}}=\frac{\tau_{_0}}{2}\sqrt{\frac{\pi}{\sigma^3}}\hspace{1mm}e^{\sigma} \hspace{3mm} \mathrm{with} \hspace{3mm} \tau_{_0}=\frac{\mu}{2\gamma kT}\frac{(1+\alpha^2)}{\alpha} \end{equation}
 controlled by the ratio of anisotropic to thermal energy $\sigma=E_K/kT$ with Boltzmann's constant $k$ and temperature $T$. At room temperature and typical saturation magnetizations the magnitude of $\tau_{_0}$ is on the order of the usual quoted value of $10^{-10}$s.

A sample of MNPs will attempt to align to a magnetic field. If a stronger field is imposed, the particles will align faster \cite{DEIS}. Only equilibrium fluctuations are considered in the N\'{e}el time quoted above, so it does not describe this phenomenon. Brown wrote a field-inclusive characteristic time from a high barrier approximation ($\sigma\gg1$) to the Fokker-Planck equation describing the thermally assisted movements between anisotropic and magnetic field energy minima\cite{WFB}
\begin{equation}\tau_{_\mathrm{hi}}=\left(\frac{\tau_{_\mathrm{N}}}{1-\epsilon^2}\right)\left(\frac{e^{\sigma\epsilon^2}}{\cosh\xi+\epsilon\sinh\xi}\right). \label{hi}\end{equation}

The applied magnetic field is accounted for with the unitless $\xi=\mu H/kT$. We define the ratio of anisotropic to magnetic energy as $\epsilon=\xi/2\sigma$ and thus when $\xi=0$, the expression reduces to the equilibrium N\'{e}el time.

From the LLG equation, it is possible to approximate the \emph{mean} magnetization if the correlation functions between variables are approximately zero. This is physically equivalent to requiring high amplitude fields or low temperatures so that the stochastic term is negligible. We are only interested in the average magnetization in the direction parallel to a constant applied field $M_z$, so we set $\mathbf{H}=\hat{z}H_z$ and the LLG equation simplifies to
\begin{equation} \frac{\mathrm{d}M_z}{\mathrm{d}t}= -\frac{\gamma\alpha H_z}{\left(1+\alpha^2\right)}\left(1-M_z^2\right) \end{equation}
where we have used the fact that the magnetization magnitude is conserved ($\mathbf{m}^2=1$). Integrating both sides assuming the initial magnetization and time are both zero results in
\begin{equation} M_\mathrm{eq}= \tanh\left[-\frac{\gamma\alpha H_z}{\left(1+\alpha^2\right)}t\right] = \tanh\left[-\frac{t}{\tau_{_c}}\right] \end{equation}
where we have interpreted the constant which has dimensions of frequency as an inverse of the characteristic time
\begin{equation}\tau_{_c} = \frac{\left(1+\alpha^2\right)}{\gamma\alpha H_z}=\frac{2\tau_{_0}}{\xi}.\label{tc}\end{equation}
This expression corroborates our intuition that the characteristic time is shorter for higher field strengths. Setting $\alpha=1$ gives the minimum $\partial_\alpha\tau_{_c}=0$ corresponding to the magnetization switching in a single precession time. For $\alpha \gg 1$ the rotations are overdamped, resulting in phase lagging. For $\alpha \ll 1$ the magnetizations will precess significantly. From this microscopic interpretation, $\alpha$ is the key to delivering heat. Unfortunately, $\alpha$ is not well understood in terms of fundamental nanoparticle variables.

The analogous derivation with an oscillating magnetic field (with $\mathbf{H}=H_o\hat{z}\cos{\omega t}$) leads to \begin{equation} M_z=-\tanh\left[\frac{\xi}{\omega\tau_{_0}}\sin\omega t\right]=-\tanh\frac{\sin\omega t}{\omega\tau_{_c}}\label{tanh} \end{equation} where now instead of a decaying magnetization, the oscillations are parameterized by a constant that is proportional to the field strength divided by the frequency. Accordingly saturation decreases with increasing frequency, reducing net relaxation losses.

We have described several ways to approximate the characteristic time of a nanoparticle sample. Now to test these times and their respective ranges of validity, we resort to numerical simulations of the LLG equation using a second order Heun integration scheme in the sense of It\={o} \cite{GARD}. The white noise field is implemented as a Wiener process with a Gaussian distribution of magnitudes. Thus statistical moments of the sample magnetization can be developed. Unless specified, each simulation uses $i=10^5$ repeated integrations (understood by using $i$ to be the number of particles) and $2^{10}$ time-steps. Nanoparticles are spherical with 5nm radii, $\alpha=1$, $M_s$=31emu/g and $T=300$K.

\begin{figure}[h!]\begin{centering}\includegraphics[width=3.5in]{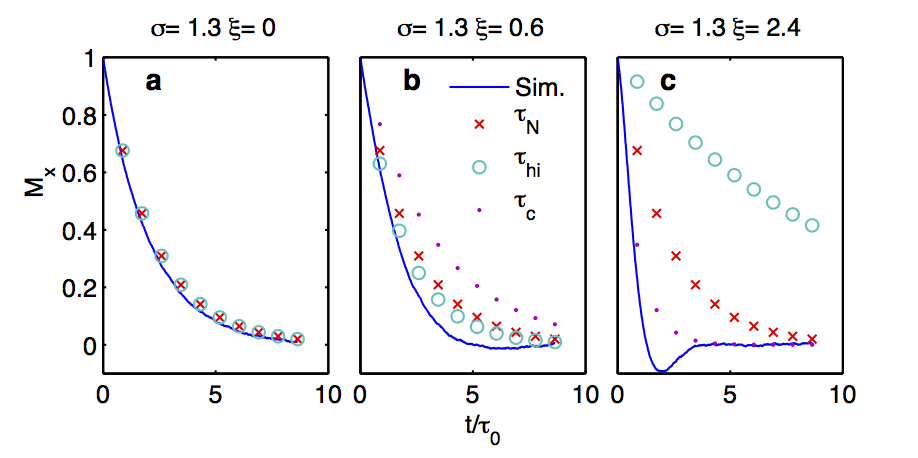} \caption{Plots of simulated magnetizations ($i=10^4$ particles) and calculated exponentially decaying magnetizations with multiple time constants based on different approximations. The combinations of $\xi$ and $\sigma$ specify the regime and the approximations match as they should.} \label{rc}  \end{centering}\end{figure}

To compare relaxation times, particles are initialized in the $\hat{x}$ direction. A constant field is instantaneously introduced in the $\hat{z}$ direction. Though the average magnetization $M_x$ decays to zero without a field due to thermal randomizing, it is forced to zero as the particles align with the field in a time determined by the field amplitude. We plot approximate average magnetizations of the form $M_x=\exp(-t/\tau)$ where the $\tau$ are our various time constants. The data for various field amplitudes and anisotropy constants are shown in Fig.~\ref{rc}. 

In Fig.~\ref{rc}(a) there is no static field, $\tau_{_\mathrm{hi}}$ is identical to $\tau_{_\mathrm{N}}$, and both agree with the stochastic simulations. Here, $\tau_{_c}$ is ill-defined. As the field increases but does not overcome the anisotropy energy, as in (b), the equilibrium expression $\tau_{_\mathrm{N}}$ is no longer accurate and $\tau_{_\mathrm{hi}}$ is the best approximation. In this regime, classic Stoner-Wohlfarth\cite{STOWO} type hysteresis curves are found. When the static field amplitude is increased beyond that of the anisotropy as in (c), the high-barrier $\tau_{_\mathrm{hi}}$ approximation breaks down because there are no longer two energy minima. At this point, $\tau_{_c}$ is the most valid approximation. Interestingly, the simulated magnetization dips below zero because the high amplitude field also causes increased precession, a physical phenomena that cannot be modeled with the simple exponential-decay model. 

Another point is that a typical MNP concentration for hyperthermia ($c_N\approx10^{13}$ particles/mL) leads to an approximate distance apart of 100nm, so that the mean dipole fields are orders of magnitude smaller than the typical hyperthermia fields, and affect the dynamics minimally. Other studies show that dipole effects are actually detrimental to heating unless obvious particle chains are formed \cite{HAASE}.
 
\begin{figure}[h!]\begin{centering}\includegraphics[width=3.25in]{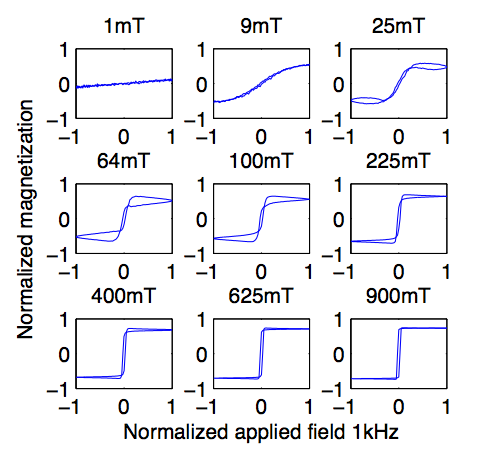} \caption{Relaxational hysteresis curves neglecting anisotropy ($\sigma=0$) for various field amplitudes at 1kHz. $\alpha =10$ is used so that the peak in loop area is visible by eye.} \label{hysteresis}  \end{centering}\end{figure}

Now simulating an oscillating applied field, we examine how the field amplitude and frequency affect the magnetization. Shown in Fig.~\ref{hysteresis}, with no anisotropy or dipole fields, we see that a hysteresis emerges in plots of the oscillating field against the resulting magnetizations and that there is a peak in area as field amplitude is increased. This can be interpreted by using the characteristic time as in Eq.~\ref{tc}. When the oscillatory applied field is at its maximum value in its cycle, the characteristic time is the shortest. Then, as the field approaches zero, the alignment is slower. Hence, the magnetization takes longer to return to zero than to saturate, and a phase-lag occurs as a result of relaxation. This is distinct from the adiabatic hysteresis that derives from the Stoner-Wolhfarth model \cite{STOWO}.

\begin{figure}[h!]\begin{centering}\includegraphics[width=3.25in]{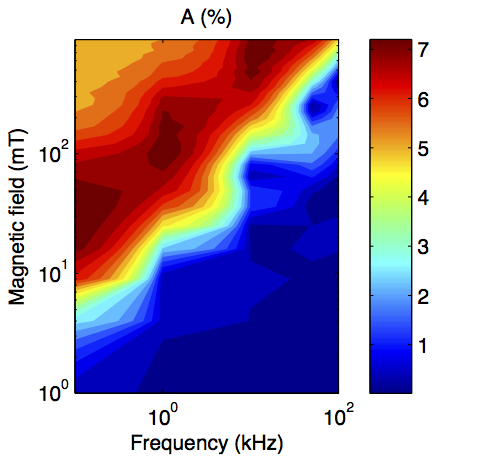} \caption{Visualization of the field and frequency dependence of the percent loop area $A$, as predicted by the magnetization dependence.} \label{contour}  \end{centering}\end{figure}

We use $A$ to denote the percentage of the total possible area covered by the normalized hysteresis loop, per cycle of the applied field. $A$ is plotted as a function of field amplitude and frequency as in Fig.~\ref{contour}. A maximum $A$ appears for certain field and frequency combinations, and has the correct qualitative scaling behavior expected theoretically by Eq.~\ref{tanh}. In particular, if the frequency is increased, the field must also be increased to maintain the same magnetization dynamics. We purposefully stay within the physically tolerable regime \cite{BREZ} of maximal magnetic field power to suggest hyperthermia could benefit from these peaks or from the flexibility to adjust field \emph{or} frequency to maintain the peak values.

The hysteresis loop area is an enticing metric to visualize peaks when simulating hyperthermia, but, to be practically important, this factor must be put in common experimental units of specific power loss per mass `SLP' of particles (W/g). Thus \begin{equation}\mathrm{SLP}=\frac{\mu H_o f}{\rho V}A\end{equation} where $\rho$ is the particle mass-density in g/m$^3$.

This definition of SLP means that increasing field-amplitudes or frequencies will increase heating. For the peaks in $A$ to affect SLP, they must overpower this linear increase. We only observe this more than linear peaking when $\alpha>1$. Because $\alpha$ acts as the strength of the rotational magnetic viscosity (the constant in front of the velocity dependent drag), it should be calculated from nanoparticle properties, yet for now remains an experimentally determined parameter and must be treated with care.

\begin{figure}[h!]\begin{centering}\includegraphics[width=3.25in]{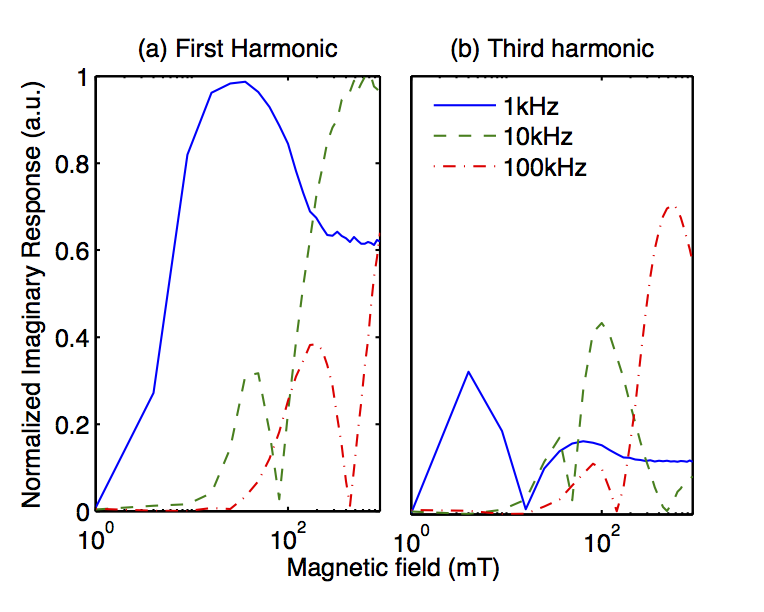} \caption{The first and third imaginary harmonic components of a Fourier transform of the magnetization. Peaks at specific field amplitudes and frequencies are visible, and significant higher harmonics are highlighted.} \label{imag}  \end{centering}\end{figure}

Another metric for the dissipative losses is the imaginary component of a Fourier transform of the magnetization. From the same numerical simulations, the imaginary first and third harmonics at several frequencies are plotted in Fig.~\ref{imag} for a range of field amplitudes. The data are normalized to the maximum value of the first harmonic. In Fig.~\ref{imag}(a) the same peaks are visible at the same fields and frequencies as compared to Fig.~\ref{contour}. In (b) the third harmonic response indicates significant nonlinear components in the lagging magnetization and peaked structure as well. Clearly, here, a physical description only including linear response theory would be incomplete to model heating.

We have demonstrated stochastic methods that allow generalized hyperthermia modeling throughout regimes including equilibrium, large anisotropy, and large applied field amplitudes. We specifically examine the regimes where each approximation is valid to elucidate regime-appropriate timescales. We show in particular that the equilibrium relaxation times will not in general describe the timescales for particle rotations in magnetic fields and that linear response theory must be used with care as higher harmonics change the physics substantially and the SLP is not reducible to a simple power law \cite{HERGT}. 

We find that a lagging mechanism that causes heat deposition (due to relaxation effects) contains field amplitude and frequency peaks that can be visualized in the imaginary components of the Fourier transform as well as via the hysteresis loop area. Varying the field or the frequency accordingly keeps the particles in the optimal regime, a result that can be qualitatively seen from the analytic solution to the high-field magnetization (Eq.~\ref{tanh}). Adding anisotropic contributions will increase heating when the applied field is small enough so that there are still two minima (see Fig.~\ref{rc}). In this case, the relaxation time can be computed from the high barrier approximation Eq.~\ref{hi}. The present results suggest the possibility for tuning the field and frequency separately to optimize heating while maintaining realistic power ranges that are physiologically relevant and practically engineered. Though we only consider N\'{e}el rotations in this work, we expect a similar relaxational hysteresis for particles that physically rotate given the similar form of the Brownian differential magnetization equation. In either case, any advantages deriving from the peaks in relaxational hysteresis could be used concurrently with the significant advances garnered from engineering the particles themselves.

\begin{acknowledgments}
The authors gratefully acknowledge the William H. Neukom 1964 Institute for Computational Science and the support of NIH-NCI Grant No.1U54CA151662-01.
\end{acknowledgments}

\bibliographystyle{unsrt}

\end{document}